# X-ray imaging of chemically active valence electrons during a pericyclic reaction

Timm Bredtmann[1], Misha Ivanov[1,2] & Gopal Dixit[1]

[1]Max Born Institute, Max-Born-Strasse 2A, 12489 Berlin, Germany.

[2]Blackett Laboratory, Imperial College London, London SW7 2AZ, United Kingdom.

**Time-resolved imaging of chemically active valence electron densities is a long sought goal, as these electrons dictate the course of chemical reactions. However, x-ray scattering is always dominated by the core and inert valence electrons, making time-resolved x-ray imaging of chemically active valence electron densities extremely challenging. To image such electron densities, we demonstrate an effective and robust method, which emphasizes the information encoded in weakly scattered photons. The degenerate Cope rearrangement of semibullvalene, a pericyclic reaction, is used as an example to visually illustrate our approach. Our work also provides experimental access to the long-standing problem of synchronous *versus* asynchronous bond formation and breaking during pericyclic reactions.**

For over a century, x-ray and electron scattering have been indispensable in studying the structure of matter with atomic-scale spatial resolution. Thanks to enormous technological progress, it is now becoming possible to generate tunable, intense, ultrashort x-ray[1–3] and electron pulses[4,5], adding femtosecond temporal resolution to structural analysis. These pulses promise to provide time-resolved snapshots of physical, chemical and biological processes in individual molecules[6–8]. Atoms in molecules are glued together by valence electrons, which undergo ultrafast rearrangements during the formation and breaking of chemical bonds. Thus, the ability to follow the flow of valence electron density is paramount for '*filming their motion*', and, ultimately, better understanding and controlling chemical reactions. The remarkable properties of modern ultrashort x-ray and electron pulses seem to offer a natural way for extending static scattering techniques into the domain of ultrafast electronic processes.

However, x-ray scattering from valence electrons is very weak in comparison to that from core electrons. The valence electron density contributes only a very small fraction to the total scattering pattern. Moreover, typically only a small fraction of the valence electron density participates actively in chemical reactions. These factors make the use of x-ray scattering for time-resolved imaging of changes in chemically relevant parts of the valence electron density during complex chemical reactions extremely challenging. One way to circumvent the low sensitivity of x-ray scattering to valence electrons, proposed over four decades ago, is to use nonlinear x-ray scattering[9]. In a recent experiment, sum and difference frequency generation in the x-ray domain have been used to determine the valence electron density of diamond in real space, for a particular orientation[10]. However, realizing the time-resolved version of nonlinear x-ray scattering requires the triad of two optical and one ultrashort x-ray pulse with controlled variable time-delays between all three, by no means an easy task. Another option is to focus on the temporal behavior of Bragg peaks. This approach has been successfully applied to imaging valence electron density in Refs. 6, 11, 12. However, it is limited to crystals, and requires that their static structure is known in advance, before the dynamical process is initiated.

Here we describe an effective and robust approach which allows us to extract the changes in the valence electron density from the overall x-ray scattering pattern dominated by the core electrons. This allows us to image the flow of valence electrons in space and time during a chemical reaction and thus solve the problem which has hampered the progress of time-resolved imaging of chemical reactions. Our approach works in both the condensed and the gas phase, giving access to valence electron rearrangements in individual molecules. We illustrate our approach using the example of a very general pericyclic reaction – the degenerate Cope rearrangement of semibullvalene, sketched in Figure 1. Our example also addresses another important and long-debated issue: is the breaking of old bonds during pericyclic reactions synchronous with the formation of new bonds? We show that our method distinguishes these two alternatives.

Figure 1A shows effective reaction paths for the degenerate Cope rearrangement. In general, the reaction path strongly depends on the preparation of the reactants. Depending on the reactant energy, the Cope rearrangement can proceed by tunneling (path 1) or over the barrier (path 2).

Such alternative pathways are ubiquitous in many chemical reactions. They invariably trigger the question about synchronous *versus* asynchronous bond breaking and formation, debated over many decades for such pericyclic reactions[13–19]. So far, the debate has been purely theoretical.

Recent quantum mechanical *ab-initio* calculations for the Cope rearrangement of semibullvalene in the electronic ground state predicted that the synchronicity of the underlying electronic bond-to-bond fluxes depends on the energy (temperature) of the reactant[16,17]. Synchronous bond formation and breaking was predicted for tunneling, which is the dominant pathway at low energies (temperatures). In contrast, asynchronous bond breaking and formation was predicted for the over the barrier pathway, typical for high excitation energies (high temperatures). In this case, there should be a time-delay between the breaking of the old and the formation of the new bonds. Initiating pericyclic reactions at different excitation energies is possible using ultrashort optical pulses: down-chirped pump-dump and other methods have been proposed and experimentally demonstrated[20,21].

To simulate the time-resolved scattering pattern, we have used the following expression for the differential scattering probability[22]

$$\frac{dP}{d\Omega} = \frac{d\sigma_{\text{th}}}{d\Omega} \int dt\, j_x(t) \left\langle \chi(\mathbf{R}, \tau) \left| \left| \int d^3r\, \rho(\mathbf{r}; \mathbf{R})\, e^{i\mathbf{Q}\cdot\mathbf{r}} \right|^2 \right| \chi(\mathbf{R}, \tau) \right\rangle. \quad (1)$$

Here, $d\sigma_{\text{th}}/d\Omega$ is the Thomson scattering cross section, $j_x(t)$ is the fluence of the incident x-ray pulse at time $t$, $\chi(\mathbf{R}, \tau)$ corresponds to the nuclear wave packet with $\mathbf{R}$ denoting the set of nuclear coordinates, $\rho(\mathbf{r}; \mathbf{R})$ is the electronic density with $\mathbf{r}$ as electronic coordinate, and $\mathbf{Q}$ is proportional to the momentum transfer of the scattered x-ray. $\rho(\mathbf{r}; \mathbf{R})$ is evaluated using quantum chemical calculations based on density functional theory using the B3LYP functional with the cc-pVTZ basis sets by means of the MOLPRO program package[23]. The time-dependent nuclear Schrödinger equation is solved to compute the time evolution of $\chi(\mathbf{R}, \tau)$ along the reaction coordinate $\xi$ describing the direct path from the reactant to the product as shown in Figure 1A. Comparison to high level *ab-initio* methods shows that (i) the structures and energies along $\xi$ are accurate in the present calculations and (ii) asynchronous bond making and breaking is a stable phenomenon, irrespective of the actual reaction path over the barrier[24]. The mean total energy along $\xi$ for the reaction over the barrier is set to 1.25 eV, well above the potential barrier of 0.36 eV, and well below the first excited electronic state which is about 4 eV higher in energy according to multi-reference configuration interaction calculations using the cc-pVTZ basis sets. Consequently, the dynamics along $\xi$ is essentially decoupled from the other degrees of freedom[25] and the reaction is well described within the Born-Oppenheimer approximation. Furthermore, the energy gap to the first excited state exceeds by far the bandwidth of the 1 eV x-ray pulses used to simulate the time-resolved scattering patterns. Under such conditions, any x-ray induced electronic transitions during the scattering process can be easily filtered out in energy-resolved x-ray scattering. Consequently, the contribution of excited electronic states and, hence, of the electronic current density[22, 26] to the scattering pattern is negligible.

Time-resolved scattering patterns in the $Q_y$ - $Q_z$ plane ($Q_x = 0$) and the corresponding electron densities of space-fixed semibullvalene in the *y* - *z* plane as a function of the delay time at times *0, T/4, T/2, 3T/4* and *T* are presented in Figure 2. Here *T* is the reaction time for the Cope rearrangement of semibullvalene, ranging from *T* = 24.2 fsec (1 fsec = $10^{15}$ sec) for the over the barrier reaction to *T* = 970 sec for the reaction via tunneling at cryogenic temperatures[27]. The static structure of semibullvalene in the gas-phase has been measured using electron scattering[28]. Figures 2A and 2C show that the time-resolved scattering patterns distinguish the over the barrier reaction from tunneling. In the tunneling case, the intensity of scattered photons, associated with high **Q**-values, diminishes from the reactant to the reaction intermediate at *T/2* and then increases again between *T/2* and *T* (see Fig. 2A). The opposite behavior is observed for the over the barrier reaction. The difference between the two reaction paths is most prominent at *T/2*. The origin of these differences is in the different rearrangement dynamics of the carbon core electrons for the two paths, which dominate the electron density as visually shown by localized yellow-green circles. The contribution of the valence electrons is very diffuse, shown by the red color in Figures 2B and 2D (for tunneling and over the barrier pathways, respectively).

Even though the full time-resolved scattering patterns distinguish tunneling from the over the barrier reaction, the changes in chemical bonding are hardly visible since the corresponding electron densities do not peak at the positions of the bonds[29–32]. We now demonstrate how the chemically active valence electron densities, which carry invaluable information about chemical reactions and hence electronic bond-to-bond fluxes, can be directly accessed from the full scattering patterns.

For the analysis of chemical bonding, topological analysis[29–32] or a partitioning of the total electronic density based on localized molecular orbitals[33,34] is typically used. In Refs. 16–18 such analyses enabled the prediction of synchronous *versus* asynchronous bond making and bond breaking, c.f. Figure 1B. Specifically, the total electronic density was partitioned into a core electron density $\rho_{core}(\mathbf{r}; \mathbf{R})$ (which accounts for the sixteen carbon core electrons), a pericyclic electron density $\rho_{peri}(\mathbf{r}; \mathbf{R})$ (which accounts for the six electrons associated with the mutation of the Lewis structure for the reactant into that of the product) and a density of the remaining valence electrons $\rho_{oval}(\mathbf{r}; \mathbf{R})$, such that $\rho(\mathbf{r}; \mathbf{R}) = \rho_{core}(\mathbf{r}; \mathbf{R}) + \rho_{peri}(\mathbf{r}; \mathbf{R}) + \rho_{oval}(\mathbf{r}; \mathbf{R})$. Although the pericyclic density has been established as a powerful theoretical tool for the analysis of pericyclic reactions as well as to access the underlying bond-to-bond fluxes, no method exists to observe this quantity experimentally.

We know that the well-localized core electrons scatter strongly and thus contribute mostly to the high **Q**-region of the scattering pattern. The delocalized valence electrons scatter weakly and show their fingerprints in the low **Q**-region of the same pattern. We take advantage of this fact by retrieving the valence electron density from the simulated scattering pattern restricted to the low **Q**-region, $\mathbf{Q}_{max} > \mathbf{Q}_{limited}$. Performing the (restricted-**Q**) Fourier transform from the **Q**-space back to coordinate space requires knowledge of the phase of the scattering pattern. Several phase retrieval algorithms are available to reconstruct the phase from the total scattering pattern[35]. Here, we have used the phase of the scattering patterns from the theoretical simulation. Thus, the proposed method combines wide-angle and small-angle scattering. The wide-angle scattering yields the complete scattering pattern necessary for phase reconstruction. The small-angle scattering pattern, combined with the phase information, is used to perform the restricted-**Q** Fourier transform. Furthermore, we take advantage of using the spatial resolution encoded in the wide- angle scattering pattern extending up to $|\mathbf{Q}_{max}| = 15$ Å$^{-1}$, which corresponds to 0.42 Å spatial resolution.

The retrieved electron densities, using the restricted **Q**-region, in the *y - z* plane are presented in Figures 3A and 3C for tunneling and the over the barrier reaction, respectively. For comparison, the pericyclic electron densities ($\rho_{peri}(\mathbf{r; R})$) for tunneling and the over the barrier reaction at times *0, T/4, T/2, 3T/4* and *T* are shown in Figures 3B and 3D, respectively. A value of $\mathbf{Q}_{limited}$ equal to 3.4 Å$^{-1}$ is used for Figures 3A and 3C, which corresponds to 1.8 Å spatial resolution, close to the characteristic radius of semibullvalene. To ensure robust reconstruction, we have varied $\mathbf{Q}_{limited}$, and find that the key features of the reconstructed densities are present for $\mathbf{Q}_{limited}$ ranging from 2 Å$^{-1}$ to 8 Å$^{-1}$.

Figure 3 shows that the hitherto dominant contribution from the core electrons is effectively filtered out, enabling a natural and tractable partitioning of the total electron densities. The reconstructed electron densities are in excellent agreement with $\rho_{peri}(\mathbf{r; R})$, which provides direct insight into the reaction mechanism. The issue of synchronous *versus* asynchronous bond formation and breakage can indeed be addressed using the reconstructed electron densities. The reconstructed electron densities for the reaction intermediate at *T/2* show that in the tunneling case, the reaction is synchronous, whereas it is indeed asynchronous in the over the barrier case. At *T/2*, the electronic flux out of the old bond, centered around $z = 1$ Å and $y = 0$ Å, is larger for the over the barrier reaction than for tunneling (disconnected and connected black contours in the vicinity of the old bond, respectively). At the same time, the flux into the new bond centered around $z = 1$ Å and $y = 0$ Å, is smaller for the over the barrier reaction as compared to the tunneling scenario.

In conclusion, using the degenerate Cope rearrangement of semibullvalene as an example, we have demonstrated a powerful approach to retrieve valence electron density from the full time- resolved scattering pattern using the restricted **Q**-reconstruction method. The phase of the scattering pattern needed for the reconstruction can be obtained from the full, unrestricted-**Q**, scattering pattern. Our approach enables us to image the instants of bond formation and breakage during chemical processes and resolve experimentally the long-standing debate of synchronous *versus* asynchronous bond formation and breakage during chemical reactions. Of course, our approach also applies for imaging chemically active valence electron densities in static structures and non-adiabatic chemical reactions[36]. It is general and applicable not only to ultrafast time-resolved x-ray scattering but also to electron scattering, which are emerging as the methods of choice for imaging ultrafast chemical and biological processes with atomic-scale spatial and temporal resolutions. The feasibility of single molecule imaging in the gas phase using x-ray[37] and electron[38] scattering has recently been experimentally demonstrated, which provides the first frame of time-resolved chemical processes. Three-dimensional molecular alignment is also experimentally feasible[39], but combining such alignment with x-ray imaging is a major challenge. The problem may be circumvented by using Van der Waals crystals, which will also greatly enhance the required signal.

The notion of quantum fluxes along with their respective densities played a pivotal role for gaining insight about complex chemical reactions. However, the concept of the fluxes and their densities has so far been limited to theoretical modeling. Our reconstruction allows us to visualize (probe) the fluxes and their densities. The imaged electron density can be directly used to determine bond-to-bond electronic fluxes during chemical reactions, essential for the deeper understanding of complex chemical reactions. Their direct imaging opens new possibilities for manipulating and controlling chemical processes.

**Acknowledgements** We thank Jochen Kuepper, Joern Manz, Jon Marangos, Arnaud Rouzee, Robin Santra and Michael Woerner for useful discussions.

Competing Interests The authors declare that they have no competing financial interests.

**Correspondence** Correspondence and requests for materials should be addressed to Timm Bredtmann and Gopal Dixit (email: Timm.Bredtmann@mbi-berlin.de, dixit@mbi-berlin.de).

**Figure 1** Degenerate Cope rearrangement of semibullvalene: (A) A cut through the potential energy surface along the effective reaction coordinate $\xi$ from the reactant to the product including a sketch of the investigated reaction paths: tunneling (green arrow), and over the barrier reaction (blue arrow). (B) Lewis structures of semibullvalene with the pincer-shaped arrows symbolizing electronic fluxes associated with chemical bond formation and breakage. The time sequences of these fluxes are sketched by the scale of the associated arrows (bonds associated with carbon atoms C4-C6 and C8-C2 are made and broken, respectively; double bonds change perpetually from C3=C4 and C6=C7 to C2=C3 and C7=C8). Synchronous bond formation and breakage is predicted for tunneling, whereas, for the over the barrier reaction, asynchronous electronic rearrangement with dominant electronic fluxes out of the old bond from the reactant to the barrier, and reversed flux patterns from the barrier to the product are predicted.

**Figure 2** Time-resolved scattering patterns in the $Q_y$ - $Q_z$ plane ($Q_x = 0$) and corresponding electron densities in the $y$ - $z$ plane (integrated along $x$ direction) during the Cope rearrangement of semibullvalene at pump-probe delay times $0, T/4, T/2, 3T/4$ and $T$. (A) Scattering patterns, and (B) electron densities for the reaction via tunneling. (C) Scattering patterns, and (D) electron densities for the reaction over the barrier. The time-resolved patterns are calculated for $|\mathbf{Q}_{max}| = 15$ Å$^{-1}$ and the grid spacing in **Q**-space is 0.17 Å$^{-1}$. The x-ray propagation direction is along the x-axis in the present case. The intensities of the patterns are shown in units of the differential scattering probability, $dP_e/d\Omega$, in both cases with $\frac{dP_e}{d\Omega} = \frac{d\sigma_{th}}{d\Omega} \int dt\, j_x(t)$.

**Figure 3** Reconstructed valence electron densities obtained via the restricted **Q**-reconstruction method and the pericyclic electron densities ($\rho_{peri}(\mathbf{r}; \mathbf{R})$) obtained via partitioning the total electron densities in the *y - z* plane for the Cope rearrangement of semibullvalene at pump - probe delay times *0, T/4, T/2, 3T/4* and *T*. (A) Reconstructed valence electron densities, and (B) $\rho_{peri}(\mathbf{r}; \mathbf{R})$ for the reaction via tunneling. (C) Reconstructed valence electron densities, and (D) $\rho_{peri}(\mathbf{r}; \mathbf{R})$ for the over the barrier reaction. $\rho_{peri}(\mathbf{r}; \mathbf{R})$ accounts for the six rearranging valence electrons according to the Lewis structures (cf. Fig. 1B). The reconstruction of the densities is performed using scattering intensities information up to $|\mathbf{Q}_{limited}| = 3.4$ Å$^{-1}$ from the full time-resolved scattering patterns as shown in Figure 2.

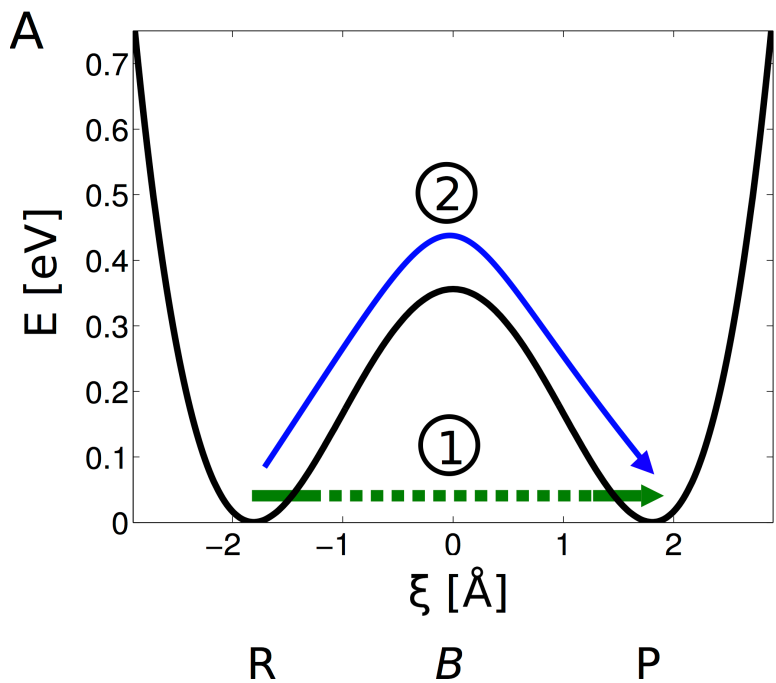

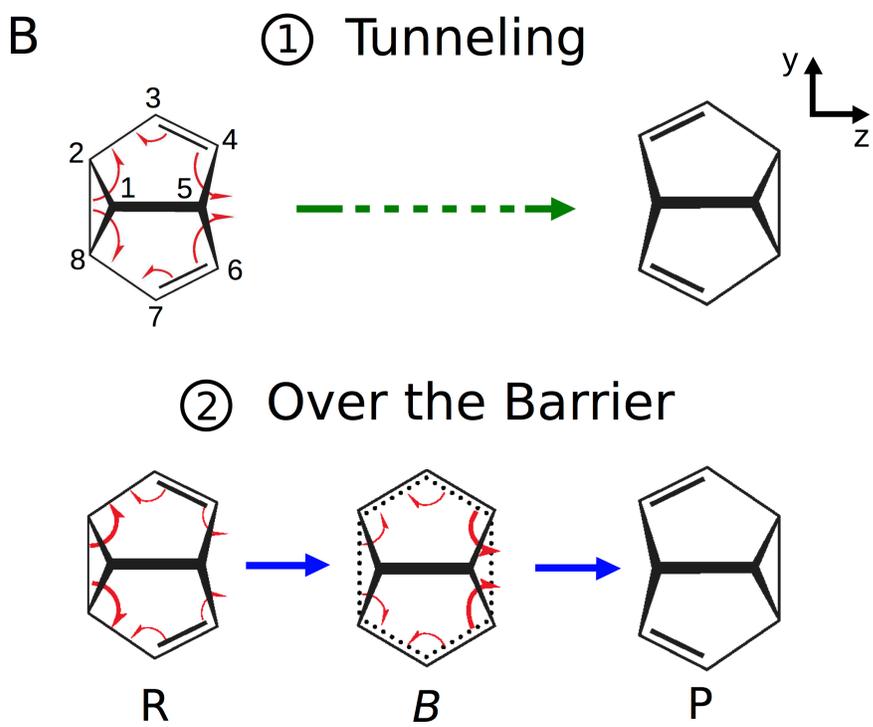

Bredtmann, Ivanov and Dixit

Figure 1

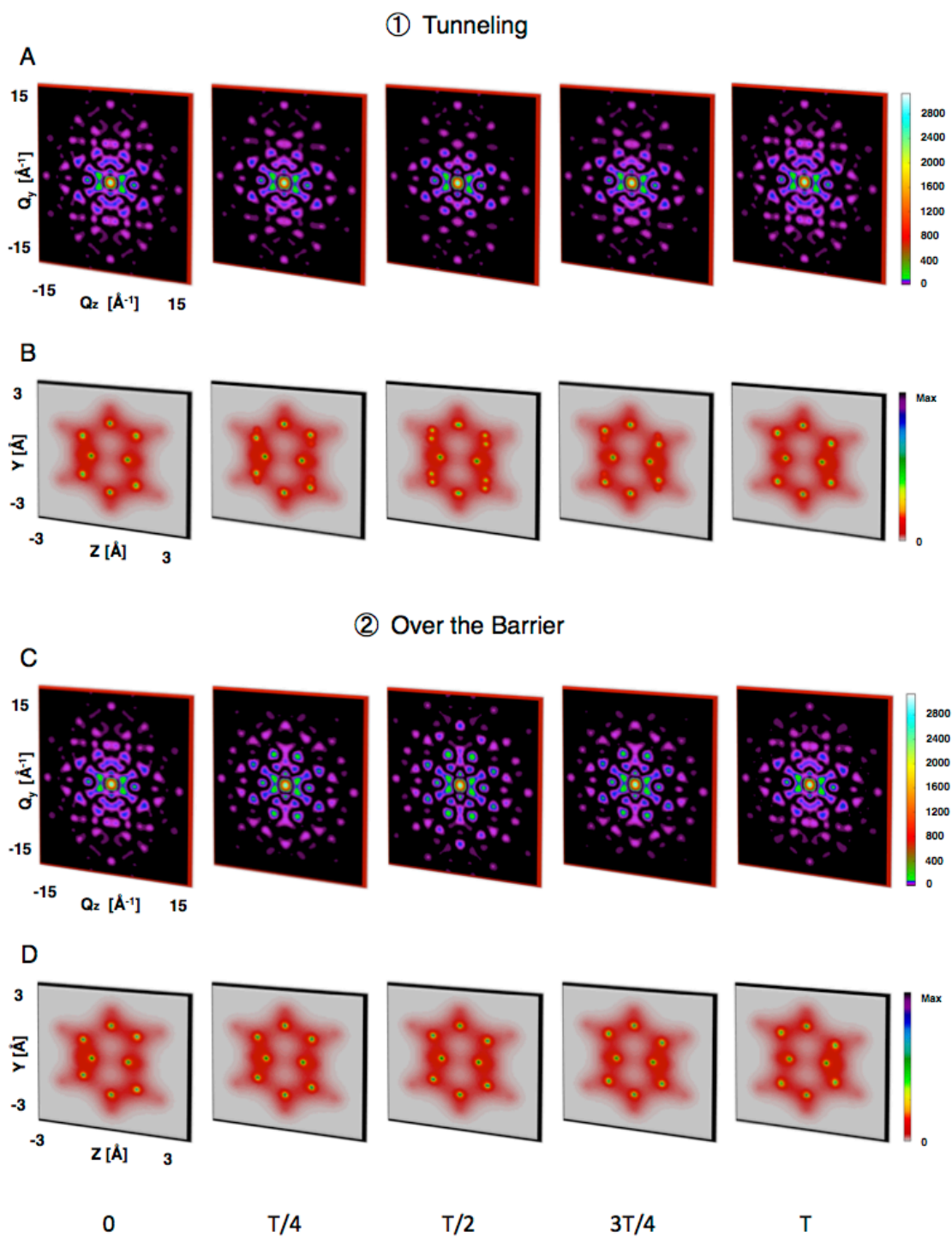

Bredtmann, Ivanov and Dixit

Figure 2

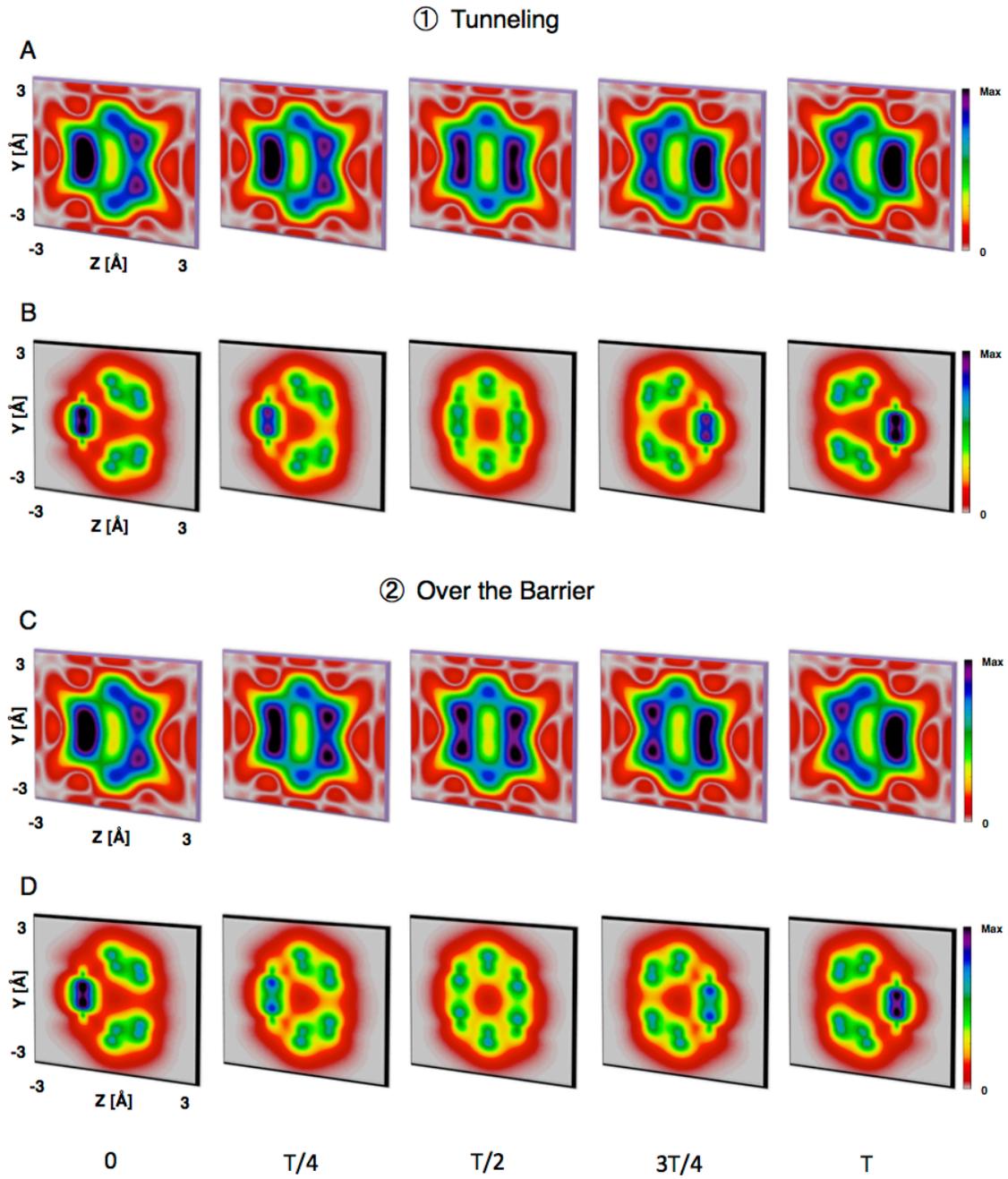

Bredtmann, Ivanov and Dixit

Figure 3